\documentclass[conference]{IEEEtran}

\ifCLASSINFOpdf
  \usepackage[pdftex]{graphicx}
  \graphicspath{{../pdf/}{../jpeg/}}
  \DeclareGraphicsExtensions{.pdf,.jpeg,.png}
\else
  \usepackage[dvips]{graphicx}
  \graphicspath{{../eps/}}
  \DeclareGraphicsExtensions{.eps}
\fi
\usepackage{cite}
\usepackage[cmex10]{amsmath}
\usepackage{amsthm, mathtools}
\usepackage{amssymb}
\usepackage{algorithm}
\usepackage{algorithmic}
\usepackage{caption}
\usepackage{subcaption}
\usepackage{float}
\usepackage{tikz,pgfplots}
\definecolor{darkgreen}{rgb}{0.12549019607843137255,0.4980392156862745098,0.16862745098039215686}

\usepackage[figwidth=\columnwidth]{todonotes}

\usetikzlibrary{arrows,calc,shadows,positioning,shapes.geometric,decorations.shapes,decorations.markings,patterns}

\IEEEoverridecommandlockouts

\begin{document}
%
\title{Characterization of Coded Random Access with Compressive
  Sensing based Multi-User Detection}

\author{\IEEEauthorblockN{Yalei Ji$^{(1)}$, \v Cedomir Stefanovi\' c$^{(2)}$, Carsten Bockelmann$^{(1)}$, Armin Dekorsy$^{(1)}$, Petar Popovski$^{(2)}$} 
 \IEEEauthorblockA{$^{(1)}$Department of Communications Engineering
  University of Bremen, Germany\\
  \{ji, bockelmann, dekorsy\}@ant.uni-bremen.de}
	\IEEEauthorblockA{$^{(2)}$Department of Electronic Systems
	Aalborg University, Denmark\\
	\{cs, petarp\}@es.aau.dk}
}


%

\maketitle

\begin{abstract}
The emergence of Machine-to-Machine (M2M) communication requires new Medium Access Control (MAC) schemes and physical (PHY)
layer concepts to support a massive number of access requests. The concept of coded random access, introduced recently, greatly outperforms other random access methods and is inherently capable to take advantage of the capture effect from the PHY layer. Furthermore, at the PHY layer, compressive sensing based multi-user detection (CS-MUD) is a novel technique that exploits sparsity in multi-user detection to achieve a joint activity and data detection. In this paper, we combine coded random access with CS-MUD on the PHY layer and show very promising results for the resulting protocol. 
\end{abstract}

\section{Introduction}
\label{sec:intro}

In recent years there has been a revival of the research interest in
random access protocols, instigated by the growth of
machine-to-machine (M2M) communications. This is particularly valid
for cellular networks, where the random access mechanisms are used to establish the initial connection between the
terminals and the base station (BS), and facilitate access to data services.  Cellular random access is commonly based on the traditional slotted ALOHA (SA), a
simple distributed access method that provides
satisfactory performance for human-oriented traffic.  However, the M2M
traffic has fundamentally different requirements, primarily seen in
the expected number of accessing terminals, and using traditional SA
may create bottlenecks already in the access network.

A promising approach for enhancing the performance of SA by using interference cancellation (IC) was demonstrated recently in
\cite{CGH2007,L2011,SP2013}. 
In brief, application of IC potentially
unlocks the collision slots, radically boosting the throughput of SA.
Of a particular importance is Liva's paper \cite{L2011}, where it
was shown that the use of successive IC in SA for recovering user transmissions 
is analogous to the iterative belief-propagation
(BP) erasure decoding, promoting the use of the erasure coding theory to design ``coded slotted ALOHA''
schemes.

Despite the similarities, there are important differences between erasure coding and SA with IC due to the effects of the wireless medium in the latter. When the power of one of the colliding signals is stronger than the rest,
a capture effect may occur, i.e., the corresponding transmission may
be successfully received. Therefore, the capture effect may
significantly affect both the access scheme design and performance, as the collision slots may potentially be exploited both through
captures and IC. In that case, the
analysis and design of coded SA schemes requires incorporation of the
capture effect in the model that is inspired by erasure coding. A brief treatment of the capture effect in coded SA was presented in
\cite{L2011}, introducing the general modification of the and-or tree
evaluation \cite{LMS1998}, a tool used to assess the
asymptotic performance of the erasure codes when decoded by
the iterative BP algorithm.  This analysis was extended further in
\cite{ALOHACAP}, showing how to actually evaluate the asymptotic
performance of coded SA for narrowband systems and Rayleigh fading.

Besides the work on MAC random access protocols there has also been a renewed interest in multi-user detection (MUD) in the context of random access. In classical MUD, it is assumed that the set of active users is known a priori and the focus on reliable data detection. However, in random access schemes the main ingredient is the uncertainty about the set of active users, such that both the user activity as well as data have to be estimated. Considering that the setup leads to so-called \emph{sporadic communication} the active users only constitute a small subset of all users, such that the problem is inherently sparse and motivates the use of compressive sensing (CS) to facilitate a low-overhead PHY scheme for low data rate M2M communication. This novel compressive sensing based multi-user detection (CS-MUD) achieves a joint detection of activity and data of the subset of active users in a slot and exhibits performance close to the 
genie-upper bound when the user activities are known a priori \cite{CS-MUD, ISWCS11, VTC12}.

In this paper we focus on the coded SA with capture effect in
broadband, MUD systems, which, to the best of
our knowledge, has not been analyzed in the literature available so
far. First, we show how to generally model and incorporate the capture
effect in MUD systems into the and-tree
evaluation, applicable to any coded slotted ALOHA scheme.
In the next step, we deviate from the simple PHY-layer, commonly used in ALOHA schemes, and introduce the details of the receiver based on CS-MUD.
Finally, we apply the obtained analytical and numerical results to the
frameless ALOHA, a simple but effective variant of coded SA
\cite{SPV2012,SP2013}, and demonstrate how the capture effect and
features of the CS-MUD impact the design and performance of the
scheme.

The paper is organized as follows. Section~\ref{sec:background}
introduces the most important concepts of coded SA and capture effect.
Section~\ref{sec:sysmod} elaborates the system model.  The analysis of
the proposed access method is performed in Section~\ref{sec:analysis}.
Section~\ref{sec:performance} presents the asymptotic performance of
frameless ALOHA with CS-MUD.  Finally, the paper is concluded in
Section~\ref{sec:conclusion}.

\section{Background}
\label{sec:background}

\subsection{Coded Slotted ALOHA}

\begin{figure}
 \begin{center}
 \includegraphics[width=0.8\columnwidth]{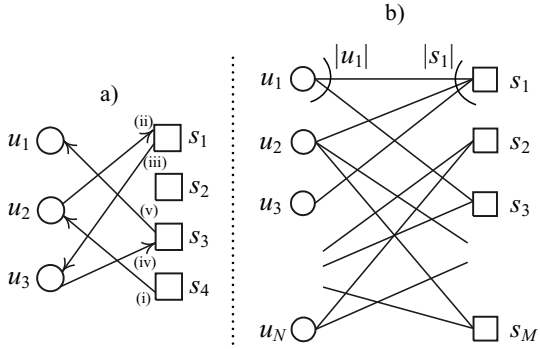}
 \caption{\small Graph representation of coded slotted ALOHA.}
 \label{fig:CSA}
 \end{center}
 \vspace{-6mm}
\end{figure}

The basic principles of coded SA are illustrated in the example presented in Fig.~\ref{fig:CSA}a).
The contention process is depicted by a bipartite graph, where nodes on the left represent users (i.e., contending terminals), the nodes on the right represent slots, and the edges connect users with the slots in which they transmitted (for instance, user $u_2$ transmitted in slots $s_1$ and $s_4$). Each time the user makes a transmission, it is the replica of the same data packet that, in addition, contains the information in which slots the other replicas occurred\footnote{A practical way how to achieve this is elaborated in \cite{SP2013}.}.
In traditional SA with no capture effect, only the singleton slot $s_4$ is usable and the packet of $u_2$ becomes recovered, denoted by (i) in Fig.~\ref{fig:CSA}a); the collision slots $s_1$ and $s_3$ and the idle slot $s_2$ are wasted.
On the other hand, the use of interference cancellation allows for removal of packet replica of $u_2$ from $s_1$ (ii), which reduces $s_1$ to a singleton slot and enables recovery of the packet of $u_3$ (iii).
In the same way, the packet replica of $u_3$ is removed from $s_3$ using IC (iv), enabling recovery of the packet of $u_1$ (v).
The above representation and the successive application of IC are analogous to the representation of codes-on-graphs and iterative BP erasure-decoding.

We now briefly introduce the notation used in the rest of the paper.
Denote users by $u_i$, $1 \leq i \leq N$, and slots by $s_j$, $1 \leq j \leq M$, see Fig.~\ref{fig:CSA}b).
By $|u_i|$ ($|s_j|$) denote the degree of $u_i$ ($s_j$), where the degree is the number of edges incident to the node.\footnote{Henceforth, we use notions of user/user node, slot/slot node and transmission/edge interchangeably.}
Further, denote by $\Lambda_{i,k}$ ($ \Omega_{j,l}$) the probability that a degree of user (slot) node is $k$ ($l$).
An important feature of coded SA is that one can design only the user degrees $|u_i|$, $1 \leq i \leq N$, while the slot degrees $|s_j|$, $1 \leq j \leq M$, are random, being outcomes of the contending process. 
A typical assumption is that users select slots in which they transmit with a uniform probability \cite{L2011,NP2012,SPV2012,SP2013}.
This implies that $|s_j|$, $1 \leq j \leq M$, are independent and identically distributed (i.i.d.), i.e., $\Omega_{j,l} = \Omega_l $. 
As shown in \cite{L2011,SPV2012}, the probabilities $ \Omega_l $, $l\geq 0 $, can be approximated by a Poisson distribution:
\begin{align}
\label{eq:Omega}
\Omega_l \approx \frac{\beta^l}{l!}e^{-\beta},
\end{align}
where $\beta = \mathrm{E} [ |s_j| ]$, $ 1 \leq j \leq M$, i.e., $\beta$ is the average slot degree.
The same assumption is typically used in sporadic communication setups with CS-MUD \cite{CS-MUD}, enabling the combination of the two approaches.

\subsection{Capture Effect}

As already stated in the introductory section, a capture effect occurs when
one or more user transmissions may become recovered despite the interference originating from other users.
In narrowband systems, the capture effect leads to recovery of a single user transmission, while in broadband MUD systems it may lead to recovery of more than just one transmission.

The capture effect has been extensively studied in the traditional SA framework, both in narrowband and in broadband systems; c.f. \cite{R1975,N1984,GVS1988,ZR1994,NEW2007,TheoreticalCapture}.
A typical premise is that the capture occurs for all transmissions whose signal-to-interference-plus-noise ratio (SINR) is above capture threshold $b$, where $b \geq 1$ for narrowband, and $b < 1$ for broadband systems \cite{ZR1994,NEW2007,TheoreticalCapture}.
However, the features of the CS-MUD receiver studied in this paper and presented in detail in Section~\ref{sec:cs-mud}, can not be described by such a simplistic model.
Specifically, the considered receiver exploits compressive sensing, and its performance depends on the sparsity of the input observation, as well as the correlation among users' spreading sequences and noise. We note that these dependencies pose significant analytical difficulties and we therefore 
numerically evaluate the capture probabilities for the CS-MUD receiver and the scenario of interest, as presented in Section~\ref{sec:captureprobability}.

\section{System Description}
\label{sec:sysmod}

\begin{figure}
 \centering
 \begin{tikzpicture}[font=\footnotesize]

\tikzset{
se/.style={-stealth, thick},
dots/.style={dash pattern=on 1pt off 4pt,very thick},
box/.style={rectangle, fill=white, inner sep=7pt, very thick, draw,drop shadow={fill=black, xshift=1mm, yshift=-1mm}},
hbox/.style={rectangle, color=white, fill=white, inner sep=7pt, very thick, draw,drop shadow={fill=white, xshift=1mm, yshift=-1mm}},
vbox/.style={box,rotate=90,text centered},
tbox/.style={rectangle, fill=blue!40, minimum width=0.25cm, draw},
antenna/.style={isosceles triangle,fill=black, shape border rotate=-90, inner sep=1pt},
sporadic/.style={-stealth,thick,dashed},
axis/.style={-stealth, thin},
axistick/.style={thin,minimum height=6mm,path picture={\draw[black,thin] (0,1mm) -- (0,-1mm);}},
data/.style={circle,thick,draw, inner sep=1.2pt},
basestation/.style={minimum height=1.2cm,minimum width=1cm,path picture={
\draw[black,thin] (0,-6mm) -- (0,4mm);
\draw[black,thin] (-4mm,-4.4mm) -- (0,4mm);
\draw[black,thin] (4mm,-4.4mm) -- (0,4mm);
\draw[black,thin] (4mm,-4.4mm) -- (0,-6mm) -- (-4mm,-4.4mm);
\draw[black,thin] (3.2mm,-2.7mm) -- (0,-4mm) -- (-3.2mm,-2.7mm);
\draw[black,thin] (2.4mm,-1.0mm) -- (0,-2mm) -- (-2.4mm,-1.0mm);
\draw[black,thin] (1.6mm,0.6mm) -- (0,0mm) -- (-1.6mm,0.6mm);
\draw[black,thin] (0.8mm,2.3mm) -- (0,2mm) -- (-0.8mm,2.3mm);
\draw[black] (0.2mm,5.8mm) -- (0.2mm,4.5mm) -- (0,4.5mm) -- (-0.2mm,4.5mm) -- (-0.2mm,5.8mm);
\draw[black,thin] (0,4.5mm) -- (0mm,4mm);
}},
}

 \node[basestation] (bs) at (0cm,0cm){};
 \node[below right=0cm and -1.4cm of bs, text width=5.8cm, text justified] (bt){BS};

 \node[antenna,right=1.5cm of bs.north] (node3) {};
 \coordinate[right=0.15cm of bs.north] (t3);
 \draw[sporadic] (node3) -- (t3);
 \node[data,below right=0.3cm of node3] (data3) {3};
 \draw[thick] (data3) -| (node3);
 \coordinate[above right=-0.2cm and 0.1cm of data3] (start3);
 \coordinate[right=2.75cm of start3] (stop3);
 \draw[axis] (start3) -- (stop3);
 \draw[red,fill, thick] let \p1=(start3) in (\x1,\y1) rectangle (\x1+1mm,\y1+3.5mm);
 \draw[red,fill, thick] let \p1=(start3) in (\x1+1.80cm,\y1) rectangle (\x1+1.90cm,\y1+3.5mm);
 \node[axistick, right= 0.0cm of start3] () {};
 \node[axistick, right= 0.3cm of start3] () {};
 \node[axistick, right= 0.6cm of start3] () {};
 \node[axistick, right= 0.9cm of start3] () {};
 \node[axistick, right= 1.2cm of start3] () {};
 \node[axistick, right= 1.5cm of start3] () {};
 \node[axistick, right= 1.8cm of start3] () {};
 \node[axistick, right= 2.1cm of start3] () {};
 \node[axistick, right= 2.4cm of start3] () {};
 \node[tbox, above right= 0.001cm and 0.445cm of start3] () {};
 \node[tbox, above right= 0.001cm and 0.745cm of start3] () {};
 \node[tbox, above right= 0.001cm and 2.245cm of start3] () {};

 \node[antenna,above=0.5cm of node3.north] (node2) {};
 \coordinate[above right= 0.045 and 0.135cm of bs.north] (t2);
 \draw[sporadic] (node2) -- (t2);
 \node[data,below right=0.3cm of node2] (data2) {2};
 \draw[thick] (data2) -| (node2);
 \coordinate[above right=-0.20cm and 0.1cm of data2] (start2);
 \coordinate[right=2.75cm of start2] (stop2);
 \draw[axis] (start2) -- (stop2);
 \draw[red,fill, thick] let \p1=(start2) in (\x1,\y1) rectangle (\x1+1mm,\y1+3.5mm);
 \draw[red,fill, thick] let \p1=(start2) in (\x1+1.80cm,\y1) rectangle (\x1+1.90cm,\y1+3.5mm);
\node[axistick, right= 0.0cm of start2] () {};
 \node[axistick, right= 0.3cm of start2] () {};
 \node[axistick, right= 0.6cm of start2] () {};
 \node[axistick, right= 0.9cm of start2] () {};
 \node[axistick, right= 1.2cm of start2] () {};
 \node[axistick, right= 1.5cm of start2] () {};
 \node[axistick, right= 1.8cm of start2] () {};
 \node[axistick, right= 2.1cm of start2] () {};
 \node[axistick, right= 2.4cm of start2] () {};
 \node[tbox, above right= 0.001cm and 0.14cm of start2] () {};
 \node[tbox, above right= 0.001cm and 0.745cm of start2] () {};

 \node[antenna,above=0.5cm of node2.north] (node1) {};
 \coordinate[above right=0.085cm and 0.12cm of bs.north] (t1);
 \draw[sporadic] (node1) -- (t1);
 \node[data,below right=0.3cm of node1] (data1) {1};
 \draw[thick] (data1) -| (node1.south);
 \coordinate[above right=-0.2cm and 0.1cm of data1] (start1);
 \coordinate[right=2.75cm of start1] (stop1);
 \draw[axis] (start1) -- (stop1);
 \draw[red,fill, thick] let \p1=(start1) in (\x1,\y1) rectangle (\x1+1mm,\y1+3.5mm);
 \draw[red,fill, thick] let \p1=(start1) in (\x1+1.80cm,\y1) rectangle (\x1+1.90cm,\y1+3.5mm);
\node[axistick, right= 0.0cm of start1] () {};
 \node[axistick, right= 0.3cm of start1] () {};
 \node[axistick, right= 0.6cm of start1] () {};
 \node[axistick, right= 0.9cm of start1] () {};
 \node[axistick, right= 1.2cm of start1] () {};
 \node[axistick, right= 1.5cm of start1] () {};
 \node[axistick, right= 1.8cm of start1] () {};
 \node[axistick, right= 2.1cm of start1] () {};
 \node[axistick, right= 2.4cm of start1] () {};
 \node[tbox, above right= 0.001cm and 1.0425cm of start1] () {};
 \node[tbox, above right= 0.001cm and 2.245cm of start1] () {};
 \node[above=0cm of node1] (sn){sensor nodes};

 \node[antenna,below=1.0cm of node3.south] (node4) {};
 \coordinate[below right=0.085cm and 0.12cm of bs.north] (t4);
 \draw[sporadic] (node4) -- (t4);
 \node[data,below right=0.3cm of node4] (data4) {N};
 \draw[thick] (data4) -| (node4);
 \coordinate[above right=-0.2cm and 0.1cm of data4] (start4);
 \coordinate[right=2.75cm of start4] (stop4);
 \draw[axis] (start4) -- (stop4);
 \draw[red,fill, thick] let \p1=(start4) in (\x1,\y1) rectangle (\x1+1mm,\y1+3.5mm) node[above] {Beacon};
 \draw[red,fill, thick] let \p1=(start4) in (\x1+1.80cm,\y1) rectangle (\x1+1.90cm,\y1+3.5mm);
 \node[right=0.1mm of stop4] () {\footnotesize slots};
 \node[axistick, right= 0.0cm of start4] () {};
 \node[axistick, right= 0.3cm of start4] () {};
 \node[axistick, right= 0.6cm of start4] () {};
 \node[axistick, right= 0.9cm of start4] () {};
 \node[axistick, right= 1.2cm of start4] () {};
 \node[axistick, right= 1.5cm of start4] () {};
 \node[axistick, right= 1.8cm of start4] () {};
 \node[axistick, right= 2.1cm of start4] () {};
 \node[axistick, right= 2.4cm of start4] () {};
 \node[tbox, above right= 0.001cm and 0.14cm of start4] () {};
 \node[tbox, above right= 0.001cm and 1.0425cm of start4] () {};
 \node[tbox, above right= 0.001cm and 1.945cm of start4] () {};

 \coordinate[below right=0.4cm and 1.16cm of node3.south] (dots_start);
 \coordinate[above right=0.1cm and 1.16cm of node4.north] (dots_end);
 \draw[dots] (dots_start) -- (dots_end);

\draw[red] (bs.north) ++(-45:5mm) arc (-45:90:5mm);
\draw[red] (bs.north) ++(-45:10mm) arc (-45:90:10mm);
\draw[red] (bs.north) ++(-45:15mm) arc (-45:90:15mm);

\draw[decorate,decoration=brace] let \p1=(start4) in (\x1+1.64cm,\y1-1.5mm) -- node[midway,below]{$M$} (\x1+0.13cm,\y1-1.5mm) ;
\draw[darkgreen][ dashed,thick][rotate=0] (4.575,0.4) ellipse (5pt and 45pt);
 \node (cs-mud) at (4.575,-1.4) {\textcolor{darkgreen}{\footnotesize{CS-MUD}}};
\end{tikzpicture}
 
 \caption{\small Sporadic uplink transmission of multiple devices controlled by BS beacons. Each time
   slot corresponds to one data frame. The system is assumed to be
   synchronous.}
 \label{fig:m2m}
 \vspace{-6mm}
\end{figure}
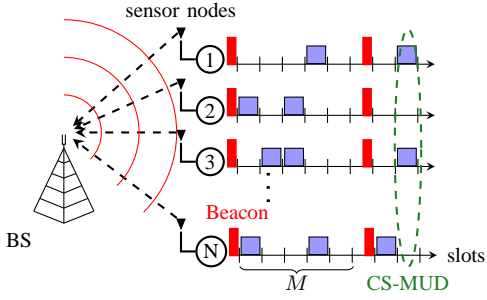

In this paper we adopt the frameless ALOHA strategy
\cite{SPV2012,SP2013}.
We assume that there are $N$ users in the system, synchronized on a
slot basis. The users contend for the access to the base station (BS)
by transmitting replicas of the same packet in randomly selected slots
of the contention period. The start and end of the contention period
are denoted by downlink beacons sent by the BS, as shown in Fig.~\ref{fig:m2m}.
The duration of the contention period in slots, denoted by $M$, is a priori unknown and chosen such that the expected throughput is maximized.
Every user
transmits with a predefined activation probability $p_A = \beta / N$ in every
slot of the contention period, where $\beta$ is a suitably chosen
constant. Therefore, the probabilities $\Lambda_{i,k}$ are the same
for all users $u_i$, and it can be shown that:
\begin{align}
\label{eq:Lambda}
\Lambda_{i,k} = \Lambda_k \approx \frac{\left( (1+\epsilon)\beta \right)^k }{k!}e^{- \beta}, \; 1\leq i \leq N,
\end{align}
where $\epsilon = \frac{M}{N} - 1 $. The average user degree is:
\begin{align}
\label{eq:avg_user_deg}
\mathrm{E}[|u_i|] = ( 1 + \epsilon)\beta, \; 1 \leq i \leq N.
\end{align}

To ease the following analytical as well as numerical investigations the following assumptions are made: (i) the user channels are assumed to be i.i.d.~and constant for the duration of a contention period, (ii) channel state information is perfectly known at the BS and (iii) the received power of all users is the same on average, e.g., through power control.

\subsection{Receiving Operation at BS}
\label{sec:BS}

In every slot $s_j$, $1 \leq j \leq M$, the BS receives and stores the
composite signal $\mathbf{y}_j$, which combines the colliding
transmissions and the noise, and executes the following steps:
\begin{enumerate}
	\item All previously recovered transmissions whose replicas occur
in $s_j$ (if there are such), are removed from $s_j$, i.e., the BS
performs the \emph{inter-slot} IC.
	\item The BS applies the CS-MUD algorithm.
	\item The BS removes all the newly recovered transmissions in step 2) from $s_j$, i.e., performs \emph{intra-slot} IC.
	\item The BS removes all the replicas of the newly recovered transmissions in step 2) from all the previous slots where they occur, i.e., performs
\emph{inter-slot} IC in $s_k$, $1 \leq k \leq j - 1$.
	\item The BS repeats steps 2)-4) until there are no newly recovered transmissions in $s_j$.
\end{enumerate}
Furthermore, steps 2)-5) are also executed on all the slots $s_k$, $1 \leq k \leq j - 1$, that experienced inter-slot IC in previous runs, potentially resulting with new candidate slots on which the same operation cycle, i.e., CS-MUD, intra- and inter-slot IC, is executed again.
The processing at the BS ultimately stops when there are no new slots affected by the inter-slot IC.

An assumption made in this paper is that both inter- and
intra-slot IC are perfect, i.e., the recovered transmissions and their
replicas are removed leaving no residual interference\footnote{In \cite{L2011} it was demonstrated that this assumption holds under moderate-to-high SNRs.}.

\subsection{Physical Layer CS-MUD}
\label{sec:cs-mud}

For a general introduction to CS-MUD in sporadic communication please refer to \cite{CS-MUD}. Here, we focus on a summary of the most important parts.
We model the physical layer transmission through a typical
simplified synchronous baseband description, in which a linear
input-output relation is used to express the multi-user wireless
transmission in slot $s_j$ \cite{Verdu}:
\begin{equation}
 \mathbf{y}_j = \mathbf{A}\mathbf{x} + \mathbf{n}_j,
\label{eq:linearequation}
\end{equation}
where $\mathbf{A}$ summarizes the influences of channel and medium
access. The stacked multi-user vector $\mathbf{x}$ contains all symbols from all $N$ user
nodes in one slot, independent of their activity in that slot. Due to the
probabilistic activity of users, i.e., the number of active users in one slot is given by the random slot degree, the activity in the system is unknown
at the BS and needs to be estimated. Accordingly, inactive users who
do not transmit any data are modeled as ``transmitting'' a frame of
$L$ zeros. Active users transmit frames of $L$ data
symbols from a discrete modulation alphabet $\mathcal{A}$. Thus, the elements of the multi-user vector are $\mathbf{x} \in
\mathcal{A}^{LN}_0$ with $\mathcal{A}_0$ being the so-called
\textit{augmented} alphabet given by $\mathcal{A}_0 = \mathcal{A}
\cup 0$. For simplicity, we assume that BPSK is applied in our
system, which leads to $\mathcal{A}_0 = \{\pm 1, 0\}$. However, this is not a general restriction and higher order modulations are easily incorporated. Finally,
$\mathbf{n}_j$ denotes additive white Gaussian noise with zero mean and
variance $\sigma^2_\mathrm{n}$ and the vector $\mathbf{y}_j$ represents the
measurements of the signal $\mathbf{x}$ available at the receiver in slot $s_j$.

In sporadic communication the basic assumption is intermittent (sporadic) user activity, which usually leads to a small number of active users compared to the total number of users. Here, the SA scheme determines how many users are accessing a slot at any given time, which may lead to less sparse systems compared to the current CS-MUD literature. Nonetheless, many elements in $\mathbf{x}$ may be zero or, more specifically, block-zero in
blocks of $L$ consecutive symbols. In CS literature this property is
called ``block-sparse'' and enables improved detection algorithms
which exploit this structure. Thus, the sparse multi-user detection problem
\eqref{eq:linearequation} can be solved by CS, which facilitates a joint activity and data detection. A very interesting
feature of this CS-MUD is that \eqref{eq:linearequation} may even be highly
under-determined. The reconstruction of $\mathbf{x}$ from the noisy
received signal $\mathbf{y}_j$ is still possible due to the sparsity of
$\mathbf{x}$.

For the remainder of the paper, CDMA will be applied as an exemplary
medium access scheme, which is quite attractive for M2M communications
due to its adaptive and flexible support of different number of
devices, as well as variable Quality of Service. Specifically, $\mathbf{A}$
describes the spreading of $\mathbf{x}$ by user-specific real-valued
PN sequences, which are assumed to be known at the BS and could serve
as a user ID. All transmitted frames are assumed to have the same length
of $F$ chips after spreading each of the $L$ symbols by a PN sequence with
spreading factor $N_\mathrm{s}$. Here, the spreading factor $N_\mathrm{s}$ determines
the resource efficiency of the PHY layer. As the slot degree $l$
is typically much smaller than $N$, we choose $N_\mathrm{s} < N$, which leads
to an overloaded CDMA system. In this case, \eqref{eq:linearequation}
is an under-determined linear equation system. Further, the matrix $\mathbf{A}$ also describes the convolution with the user-specific channel impulse responses $\mathbf{h}_i \in \mathbb{C}^{L_\mathrm{h}}$
of length of $L_\mathrm{h}$. After convolution $F^{\prime} = F + L_\mathrm{h} -1$ chips
will be received. Thus, $\mathbf{A} \in \mathbb{C}^{F^{\prime} \times NL}$ summarizes both the spreading and channel convolution.

The details of CS-MUD algorithms
are not our main focus and readers who are interested may refer to \cite{CS-MUD}.
In order to have reasonable complexity and exploit the discussed group sparsity, we
choose the well studied Group Orthogonal Matching Pursuit (GOMP) as
our detection algorithm. Usually, physical layer
algorithms are analyzed by bit or frame error rate plots over the
signal-to-noise ratio. However, here we are more interested in the
capture probability of this specific PHY layer setup for the overall
evaluation of coded random access with CS-MUD. In
Section~\ref{sec:captureprobability}, we will detail how these
probabilities are numerically obtained.

\section{Analysis}
\label{sec:analysis}

\subsection{And-or Tree Evaluation}
\label{sec:andortree}

For the general introduction to the and-or tree evaluation in the erasure coding framework, we refer the reader to \cite{LMS1998,RU2007}.
Hereafter, we focus on the most relevant aspects that are subject to the properties of the proposed access method and the CS-MUD receiver.

\begin{figure}
 \begin{center}
 \includegraphics[width=0.65\columnwidth]{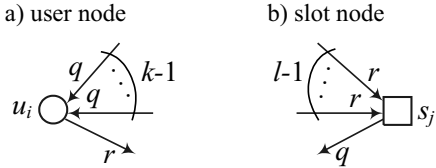}
 \caption{\small And-or tree evaluation.}
 \label{fig:probs}
 \end{center}
 \vspace{-6mm}
\end{figure}

And-or tree evaluation provides the asymptotic probability (i.e., when number of users $N \rightarrow \infty$) of user transmission recovery.
The evaluation is based on the graph that represents the contention process, Fig.~\ref{fig:CSA}b), and it is performed via iterative updates of the probabilities of transmission recoveries over the corresponding graph edges.
As elaborated in Section~\ref{sec:sysmod}, the access strategy is uniform both over users and slots, users' channels are statistically equivalent, and the expected received powers for all users are the same.
Therefore, we can model the and-or tree evaluation through a message exchanges among a referent (exemplary) user and slot, depicted in Fig.~\ref{fig:probs}.
In Fig.~\ref{fig:probs}a), $q$ denotes the probability that a user transmission corresponding to the incoming edge \emph{is not} recovered in the incident slot in the previous iteration.
Therefore, the probability $r$ that a replica corresponding to the outgoing edge \emph{is not} recovered is $r=q^{k-1}$, i.e., the outgoing edge is not recovered if all the incoming edges are not recovered.
Averaging over user degrees yields:
\begin{align}
\label{eq:r_avg}
r = \sum_{k=1}^{\infty} \lambda_k q^{k-1},
\end{align}
where $\lambda_k$ denotes the probability that an edge is connected to a user node of degree $k$, which can be calculated as \cite{RU2007}:
\begin{align}
\label{eq:lambda}
\lambda_k = \frac{k \Lambda_k}{\sum_{v=1}^{\infty} v \Lambda_v}.
\end{align}
For frameless ALOHA, using \eqref{eq:Lambda} and \eqref{eq:lambda} transforms \eqref{eq:r_avg} into:
\begin{align}
r = \sum_{k=1}^{\infty} \frac{ \left((1+\epsilon)\beta \right)^{k-1} }{(k-1)!} e^{-(1+\epsilon)\beta} q^{k-1} = e^{-(1+\epsilon)\beta(1-q)}.
\end{align}

The analysis of the probability updates performed in slots is more involved, due to the way that receiver operates.
The probability that the outgoing edge \emph{is} recovered is:
\begin{align}
\label{eq:q}
1 - q = \sum_{t=0}^{l-1} C (t) {l - 1 \choose t} (1 - r)^{l-1-t} r^{t},
\end{align}
where $t$ is the number of unrecovered incoming edges, $l-1-t$ is the number of incoming edges that have been recovered in other slots and removed via inter-slot IC, $C(t)$ is the probability that the outgoing edge is recovered when $t$ unrecovered edges remain, and ${l - 1 \choose t}$ is a combinatorial argument representing the symmetry of the problem setting.
We deal with computation of $C(t)$ in Section~\ref{sec:C(t)}.

Averaging \eqref{eq:q} over slot degrees yields:
\begin{align}
\label{eq:q_avg}
q = 1 - \sum_{l=1}^{\infty} \omega_l \sum_{t=0}^{l-1} C (t) {l - 1 \choose t} (1 - r)^{l-1-t} r^t,
\end{align}
where $\omega_l$ denotes the probability that an edge is connected to a slot of degree $l$ \cite{RU2007}:
\begin{align}
\label{eq:edge}
\omega_l = \frac{l \Omega_l}{ \sum_{v=1}^{\infty} v \Omega_v }.
\end{align}
Combining \eqref{eq:Omega}, \eqref{eq:edge} and \eqref{eq:q_avg} yields:
\begin{align}
q = 1 - \sum_{l=1}^{\infty} \frac{\beta^{l-1}e^{-\beta}}{(l-1)!} \sum_{t=0}^{l-1} C (t) {l - 1 \choose t} (1 - r)^{l-1-t} r^t.
\end{align}

Finally, the and-or tree evaluation is performed in an iterative manner:
\begin{align}
q_0 = 1 & ; \\
r_m =  f (q_{m-1}) \text{ and } q_m & =  g (r_m), \; m \geq 1,
\end{align}
where the subscript denotes the iteration number, and $f (\cdot)$ and $g (\cdot)$ are given by \eqref{eq:r_avg} and \eqref{eq:q_avg}, respectively.
The probability of user transmission recovery is:
\begin{align}
P_R = 1 - \lim_{m \rightarrow \infty} r_m.
\end{align}
A central measure of system efficiency is the throughput $T$, defined as the number of recovered users within $M$ slots of the contention period.
$T$ can be computed as:
\begin{align}
\label{eq:throughput}
T = \frac{N P_R}{M} = \frac{P_R}{1+\epsilon}.
\end{align}

\subsection{Derivation of Capture Probabilities}
\label{sec:C(t)}

Assume that $t_\mathrm{A}$ users are active in the slot, i.e., there are $t=t_\mathrm{A} - 1$ interfering transmissions from the perspective of the referent outgoing edge.
Denote by $p(s|t_\mathrm{A})$ the probability that $s$ users out of $t_\mathrm{A}$ have been recovered from the slot using CS-MUD.
Further, by $c(s|t_\mathrm{A})$ denote the probability of the event that among these $s$ is the user that corresponds to the outgoing edge.
Likewise, by $u(s|t_\mathrm{A})$ denote the event among $s$ recovered users is not the one corresponding to the outgoing edge.
Obviously:
\begin{align}\label{eq:pc(t)}
p(s|t_\mathrm{A}) = c(s|t_\mathrm{A}) + u(s|t_\mathrm{A}).
\end{align}
The recovery of the outgoing edge can happen in any chain of fortunate successive applications of the CS-MUD and the intra-slot IC, as elaborated in Section~\ref{sec:BS}, which can be stated as:
\begin{align}
\label{eq:c(t)}
C(t) = C (t_\mathrm{A} - 1) = \sum_{s_1,s_2,...,s_q} u(s_1|t_\mathrm{A})u(s_2|t_\mathrm{A} - s_1) \cdots \nonumber \\
 \cdots u \left(s_{q-1}|t_\mathrm{A} - \sum_{i=1}^{q-2} s_i \right) c \left( s_q | t_\mathrm{A} - \sum_{i=1}^{q-1} s_i \right),
\end{align}
where $q$ is the number of intra-slot IC steps and:
\begin{gather*}
 t_\mathrm{A} \geq 1, \; 1 \leq q \leq t_\mathrm{A}, \; \sum_{i=1}^{q} s_i  \leq t_\mathrm{A}, \; 1 \leq s_1 \leq t_\mathrm{A}, \\
 1 \leq s_j \leq t_\mathrm{A} - \sum_{i=1}^{j-1} s_i, \text{ for } 2 \leq j \leq q.
\end{gather*}
In other words, the pivotal idea in \eqref{eq:c(t)} is that the chain of the events, whose probabilities constitute summands, ends up with the outgoing edge being recovered.\footnote{It can
be also shown that the summands in \eqref{eq:c(t)} are mutually exclusive.}

As all the transmissions colliding in the slots are statistically equivalent, probabilities $c(s | t_\mathrm{A})$ can be obtained from $p(s | t_\mathrm{A})$ in the following way:
\begin{align}\label{eq:cst}
c(s|t_\mathrm{A}) = \frac{ { t_\mathrm{A}-1 \choose s - 1 } }{ { t_\mathrm{A} \choose s } } p ( s | t_\mathrm{A}),
\end{align}
for $t_\mathrm{A} \geq 1$ and $1 \leq s \leq t_\mathrm{A}$; $u(s|t_\mathrm{A})$ can be obtained using \eqref{eq:pc(t)}.
The analytical derivation of probabilities $p(s | t_\mathrm{A})$ for the considered CS-MUD receiver represent a formidable task that is out of the paper scope.
Instead, we obtain them numerically, as described in the next subsection.

\subsection{Numerical Evaluation of Capture Probabilities}
\label{sec:captureprobability}

We focus on the physical layer processing with CS-MUD in one slot, as described
in Section~\ref{sec:cs-mud}. In general, the performance of physical
layer algorithms is analyzed by bit or frame error rate results, which
are obtained by Monte Carlo simulations. Here, however, we define the
event of $s$ users being successfully decoded as $\Xi^{t_\mathrm{A}}_s$ in a slot
of degree $t_\mathrm{A}$. Then, the capture probability can be numerically
evaluated by extending the usual average frame error rate evaluation such that all
user frames \emph{without frame error} in the sense of $\Xi^{t_\mathrm{A}}_s$ are
counted. Evaluating $T_\text{sim}$ Monte Carlo simulations, the
capture probability can be approximated as:
\begin{equation}\label{eq:captureprobbilityapp}
  p(s|t_\mathrm{A}) \approx \frac{\#\{\Xi^{t_\mathrm{A}}_s\}}{T_\text{sim}}.
\end{equation}
where $\#\{\Xi^{t_\mathrm{A}}_s\}$ counts how many times the event has happened;
this approximation enhances with increasing $T_\text{sim}$. According to \eqref{eq:c(t)}, $p(s|t_\mathrm{A})$ is
required for all combinations of slot degrees $t_\mathrm{A} = 1, 2, \ldots,
t_\text{max}$ and captures $s=1,\ldots,t_\mathrm{A}$, where $t_\text{max}$ should
be sufficiently high to evaluate all non-zero probabilities $p(s|t_\mathrm{A})$
up to the achievable accuracy.

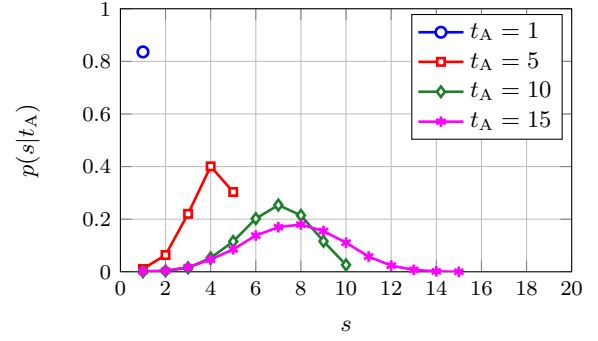
\begin{figure}
\centering
\begin{tikzpicture}

\definecolor{mycolor1}{rgb}{1,0,1}
\definecolor{darkgreen}{rgb}{0.12549019607843137255,0.4980392156862745098,0.16862745098039215686}

\begin{axis}[%
small,
view={0}{0},
width=6cm,
height=3.5cm,
scale only axis,
xmin=0, xmax=20,
xlabel={$s$},
xmajorgrids,
ymin=0, ymax=1,
ylabel={$p(s|t_\mathrm{A})$},
ymajorgrids,
legend style={nodes=right,  inner xsep=2pt, inner ysep=1pt, font=\small},label style ={font=\small}]
\addplot [
color=blue,
solid,
line width=1.0pt,
mark size=2.0pt,
mark=*,
mark options={solid,fill=white}
]
coordinates{
 (1,0.8359) 
};

\addlegendentry{$t_\mathrm{A}=1$};

\addplot [
color=red,
solid,
line width=1.0pt,
mark size=1.4pt,
mark=square*,
mark options={solid,fill=white}
]
coordinates{
 (1,0.0111)(2,0.0638)(3,0.2204)(4,0.4009)(5,0.3033) 
};

\addlegendentry{$t_\mathrm{A}=5$};

\addplot [
color=darkgreen,
solid,
line width=1.0pt,
mark size=2.0pt,
mark=diamond*,
mark options={solid,fill=white}
]
coordinates{
 (1,0.0006)(2,0.0033)(3,0.0152)(4,0.0528)(5,0.1153)(6,0.2021)(7,0.2538)(8,0.2145)(9,0.1162)(10,0.0262) 
};

\addlegendentry{$t_\mathrm{A}=10$};

\addplot [
color=mycolor1,
solid,
line width=1.0pt,
mark size=2.0pt,
mark=asterisk,
mark options={solid,fill=white}
]
coordinates{
 (1,0.0016)(2,0.0046)(3,0.0176)(4,0.0466)(5,0.0855)(6,0.1378)(7,0.1698)(8,0.1797)(9,0.1549)(10,0.1107)(11,0.058)(12,0.0236)(13,0.0075)(14,0.0018)(15,0.0001) 
};

\addlegendentry{$t_\mathrm{A}=15$};

\end{axis}
\end{tikzpicture}
\caption{\small Probability of $s$ users recovered from a slot of degree $t_\mathrm{A}$,
  for $1/\sigma^2_\mathrm{n} = 10$ dB, and $T_\text{sim} = 10^4$ simulations.}
 \label{fig:pmf_M_1_15_10dB}
\vspace{-6mm}
\end{figure}

In order to numerically evaluate the capture probability for the
and-or tree evaluation, we choose a specific physical layer
setup. As mentioned in Section~\ref{sec:sysmod}, the capture probabilities
are decided by the performance of the CS-MUD scheme. Thus, the choices of the length of PN
sequence $N_\mathrm{s}$ and average slot degree $\beta$ highly impact the numerical results. In CS theory,
the properties of $\mathbf{A}$ in \eqref{eq:linearequation} strongly determine the recovery performance. The correlation of columns in $\mathbf{A}$
is determined by PN sequences and user-specific channels and should be minimal
to achieve the best performance. However, for the sake of resource efficiency the spreading factor $N_\mathrm{s}$ should be chosen as small as possible, which requires a compromise. Finally, due to the interaction of CS-MUD and coded SA optimization in terms of average slot degrees the overall optimum choice of $N_\mathrm{s}$ is unknown and beyond the scope of this paper.

Therefore, based on the general description given in
Section~\ref{sec:cs-mud}, we focus on a basic setup to gain insight into the complex interaction of SA and CS-MUD. We assume an overloaded CDMA system
with $N = 128$ users and PN spreading sequences of length of $N_\mathrm{s} =
32$ and $L = 8$ symbols per frame. Furthermore, the transmit data will
be encoded by a convolutional code with code rate $R_\mathrm{c} = 0.5$ and
modulated to BPSK symbols. All user specific channels to the BS are
modeled as $L_\text{h} = 6$ independent and identically Rayleigh distributed taps with an
exponential decaying power delay profile. At the receiver, GOMP is
applied as the CS-MUD algorithm and $T_\text{sim}=10^4$ simulations are
performed to approximate the capture probability. All the parameters are chosen according the general system model in \cite{CS-MUD}.

\begin{figure}[t]
\centering
%
%
%
%
\begin{tikzpicture}

\begin{semilogyaxis}[%
small,
view={0}{0},
width=6cm,
height=3.5cm,
scale only axis,
xmin=0, xmax=30,
xlabel={$t$},
xmajorgrids,
ymin=0.001, ymax=1,
ylabel={$C(t)$},
ymajorgrids,
yminorgrids,
legend style={at = {(0.5,0.3)},nodes=right,  inner xsep=2pt, inner ysep=1pt, font=\small},label style ={font=\small}]
%

\addplot [
color=blue,line width=1.0pt,
solid,
mark=,
mark options={solid}
]
coordinates{
 (0,0.359)(1,0.4227877)(2,0.472600232606667)(3,0.509506361273962)(4,0.534626993159363)(5,0.56093688399171)(6,0.574700653853867)(7,0.580648846829187)(8,0.581547135824895)(9,0.571749303977967)(10,0.559299063775321)(11,0.537589120872985)(12,0.507444321203909)(13,0.46935981482181)(14,0.422505106458655)(15,0.365219509713694)(16,0.305459053461231)(17,0.24134481849225)(18,0.179145165894839)(19,0.128785068079534)(20,0.0862084313191409)(21,0.0530349705273749)(22,0.0333907788353446)(23,0.02023473994065)(24,0.0122524934053488)(25,0.00765373142336207)(26,0.00499408305251482)(27,0.00314115939644662)(28,0.00202371848133204)(29,0.00111113183662375)(30,0.000546666994037583) 
};

\addlegendentry{$1/\sigma^2_\mathrm{n} = 5 \text{ dB}$};

\addplot [
color=red,line width=1.0pt,
dashed,
mark=,
mark options={solid}
]
coordinates{
 (0,0.8359)(1,0.94385627)(2,0.96942817993)(3,0.975906932340456)(4,0.980008595050104)(5,0.982342153618381)(6,0.984493197262931)(7,0.986524159946549)(8,0.988106996666867)(9,0.989490274123879)(10,0.990619707447578)(11,0.991458534831043)(12,0.991968873292853)(13,0.992360092124568)(14,0.992788076337047)(15,0.993405428130112)(16,0.993704981059611)(17,0.993366714067072)(18,0.992461809131838)(19,0.991849725180598)(20,0.987431052865525)(21,0.977758629348576)(22,0.957498648455449)(23,0.924078480191602)(24,0.860297794028421)(25,0.760005803515146)(26,0.624427291669445)(27,0.440353985544293)(28,0.273115420103662)(29,0.13510735662508)(30,0.0492562269227092) 
};

\addlegendentry{$1/\sigma^2_\mathrm{n} = 10 \text{ dB}$};

\end{semilogyaxis}
\end{tikzpicture}
\caption{\small Capture probability $C(t)$ for and-or tree evaluation at $1/\sigma^2_\mathrm{n} = 5 \text{ and } 10$ dB.}
\label{fig:c_pmf}
 \vspace{-6mm}
\end{figure}
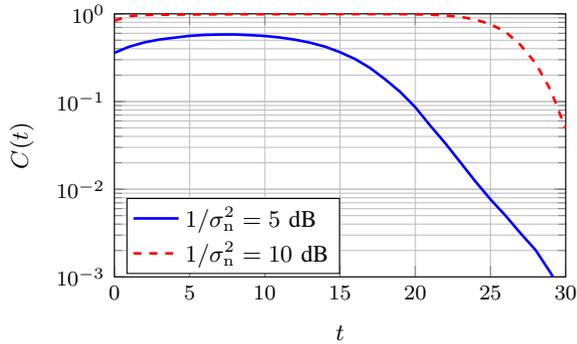

Fig.~\ref{fig:pmf_M_1_15_10dB} presents an example of the obtained probability $p(s|t_\mathrm{A})$
for $1/\sigma^2_\mathrm{n} = 10$ dB with different slot degrees
$t_\mathrm{A}$. If there is only one active user, i.e., $t_\mathrm{A} = 1$, the probability
of recovering this is $p(1|1)
\approx 0.85$. Furthermore, 
increasing $t_\mathrm{A}$ leads to lower capture probability. Particularly, the
higher $t_\mathrm{A}$ is, the lesser the sparsity of the vector $\mathbf{x}$ 
in \eqref{eq:linearequation}, which decreases the overall success
probability\footnote{The and-or tree evaluation method implicitly assumes $N\to\infty$, which cannot be evaluated with specific PHY layer processing. However, Fig.~\ref{fig:pmf_M_1_15_10dB} and  Fig.~\ref{fig:c_pmf} clearly show a strong decrease in the probability of capture such that the results are a reasonable indication of the limit performance.}.

Based on the capture probabilities provided by CS-MUD, the probability $C(t)$ for and-or tree evaluation can be obtained via \eqref{eq:c(t)}. Fig.~\ref{fig:c_pmf} presents $C(t)$ for $1/\sigma^2_\mathrm{n} = 5 \text{ and } 10$ dB. Obviously, the probabilities of a transmission recovery are higher for higher SNR.
Furthermore, the range of the number of interfering users $t$ for which the transmission recovery $C(t)$ is highly probable is significantly wider for higher SNR.

\begin{figure}[t]
    \centering
    \begin{subfigure}{\columnwidth}
    \caption{}
%
%
%
%
\begin{tikzpicture}

\begin{axis}[%
view={0}{0},
width=6.5cm,
height=2.6cm,
scale only axis,
enlarge x limits=0.05,
  try min ticks=4,
  max space between ticks=30pt,
  x tick label style={
    /pgf/number format/.cd,
    fixed,
    precision=2
  },
xmin=0, xmax=0.4,
xmajorgrids,
ymin=0, ymax=1,
ylabel={$P^\text{*}_\text{R}$},
ymajorgrids,
yminorgrids,
legend style={at = {(0.95,0.4)},nodes=right,  inner xsep=2pt, inner ysep=1pt, font=\small},label style ={font=\small}]
\addplot [
color=blue,line width=1.0pt,
solid,
mark=,
mark options={solid}
]
coordinates{
 (0.0078125,0.0447327556072206)(0.015625,0.0917599486199637)(0.0234375,0.141655940104451)(0.03125,0.19527322459494)(0.0390625,0.253835783853011)(0.046875,0.319972317918608)(0.0546875,0.398786988525928)(0.0625,0.509882351470292)(0.0703125,0.616460698996552)(0.078125,0.681995708443868)(0.0859375,0.727768641403044)(0.09375,0.764128440245958)(0.1015625,0.793582217763292)(0.109375,0.81796188007135)(0.1171875,0.837015738447631)(0.125,0.853929213988828)(0.1328125,0.86926531139006)(0.140625,0.881799949073809)(0.1484375,0.893185779318367)(0.15625,0.901744927401386)(0.1640625,0.911116821231007)(0.171875,0.919520189679746)(0.1796875,0.925774448325347)(0.1875,0.932691814086704)(0.1953125,0.938456219304029)(0.203125,0.943468964819461)(0.2109375,0.948329764252693)(0.21875,0.95237556444334)(0.2265625,0.95573723271952)(0.234375,0.959377270024054)(0.2421875,0.962895776960227)(0.25,0.965832212494185)(0.2578125,0.968376100405959)(0.265625,0.971338693033727)(0.2734375,0.973578348870376)(0.28125,0.975466845812364)(0.2890625,0.
977525559480141)(0.296875,0.979267185639458)(0.3046875,0.980959230354236)(0.3125,0.982518838249474)(0.3203125,0.983750379412981)(0.328125,0.985338510074273)(0.3359375,0.986404076662109)(0.34375,0.987734711070508)(0.3515625,0.988658265412832)(0.359375,0.98960887428455)(0.3671875,0.990591162140197)(0.375,0.991292419380674)(0.3828125,0.99205011871601)(0.390625,0.992849078293136) 
};

\addlegendentry{$1/\sigma^2_\text{n} = 5 \text{ dB}$};

\addplot [
color=black,line width=1.0pt,
dashed,
mark=,
mark options={solid}
]
coordinates{
 (0.0078125,0.152388726191322)(0.015625,0.35042978811333)(0.0234375,0.575223481457812)(0.03125,0.697307615983462)(0.0390625,0.782089971173909)(0.046875,0.843449970565449)(0.0546875,0.887654700336413)(0.0625,0.918085884794188)(0.0703125,0.939183571207874)(0.078125,0.955287853389261)(0.0859375,0.966704512491325)(0.09375,0.974497299191401)(0.1015625,0.980270014123689)(0.109375,0.985272639410225)(0.1171875,0.988630740475658)(0.125,0.991195773715331)(0.1328125,0.993386684804043)(0.140625,0.995053547184782)(0.1484375,0.996245793799063)(0.15625,0.997142097162425)(0.1640625,0.997828291847354)(0.171875,0.998449569686589)(0.1796875,0.998849579082326)(0.1875,0.999137331026633)(0.1953125,0.999359365895813)(0.203125,0.999512354265979)(0.2109375,0.999653242787508)(0.21875,0.999750020290587)(0.2265625,0.999814134466981)(0.234375,0.999861809944599)(0.2421875,0.99990194360497)(0.25,0.999930370777762)(0.2578125,0.999946379645989)(0.265625,0.999959585742483)(0.2734375,0.99997280982208)(0.28125,0.99997995560548)(0.2890625,0.
9999856775334)(0.296875,0.999988953816228)(0.3046875,0.999992095083943)(0.3125,0.99999432381468) 
};
\addlegendentry{$1/\sigma^2_\text{n} = 10 \text{ dB}$};

\end{axis}
\end{tikzpicture}\label{fig:performance-a}
        \vspace{-1mm}
    \end{subfigure}\\
    \begin{subfigure}{\columnwidth}
    \caption{}
%
%
%
%
\begin{tikzpicture}

\begin{axis}[%
view={0}{0},
width=6.5cm,
height=2.6cm,
scale only axis,
enlarge x limits=0.05,
  try min ticks=6,
  max space between ticks=30pt,
  x tick label style={
    /pgf/number format/.cd,
    fixed,
    precision=2
  },
xmin=0, xmax=0.4,
xmajorgrids,
ymin=0, ymax=25,
ylabel={$T^\text{*}$},
label style ={font=\small},
ymajorgrids,
yminorgrids]
\addplot [
color=blue,line width=1.0pt,
solid,
mark=,
mark options={solid}
]
coordinates{
 (0.0078125,5.72579271772423)(0.015625,5.87263671167768)(0.0234375,6.04398677778993)(0.03125,6.24874318703809)(0.0390625,6.49819606663708)(0.046875,6.82607611559698)(0.0546875,7.29210493304554)(0.0625,8.15811762352467)(0.0703125,8.76744105239541)(0.078125,8.72954506808151)(0.0859375,8.46858055450814)(0.09375,8.15070336262355)(0.1015625,7.81373260566934)(0.109375,7.4785086177952)(0.1171875,7.14253430141978)(0.125,6.83143371191063)(0.1328125,6.54505646223104)(0.140625,6.27057741563597)(0.1484375,6.01725156593426)(0.15625,5.77116753536887)(0.1640625,5.55347395797947)(0.171875,5.34993564904579)(0.1796875,5.15213606024541)(0.1875,4.97435634179575)(0.1953125,4.80489584283663)(0.203125,4.64477028834196)(0.2109375,4.49578554904981)(0.21875,4.3537168660267)(0.2265625,4.21842640648616)(0.234375,4.0933430187693)(0.2421875,3.97582772422287)(0.25,3.86332884997674)(0.2578125,3.75612548036251)(0.265625,3.65680449142109)(0.2734375,3.56051510444023)(0.28125,3.46832656288841)(0.2890625,3.38171004360698)(0.296875,3.
29858420425923)(0.3046875,3.21955849962416)(0.3125,3.14406028239832)(0.3203125,3.07122069670394)(0.328125,3.00293641165493)(0.3359375,2.93627260029651)(0.34375,2.87341006856875)(0.3515625,2.81218351050761)(0.359375,2.75369425887875)(0.3671875,2.69778018625415)(0.375,2.6434464516818)(0.3828125,2.5914778611357)(0.390625,2.54169364043043) 
};

\addplot [
color=black,line width=1.0pt,
dashed,
mark=,
mark options={solid}
]
coordinates{
 (0.0078125,19.5177758292619)(0.015625,22.3976807705549)(0.0234375,23.8529008103298)(0.03125,21.994221197452)(0.0390625,19.7524208299981)(0.046875,17.9130201169058)(0.0546875,16.2115240526693)(0.0625,14.7065666624536)(0.0703125,13.3017183598757)(0.078125,12.199317400885)(0.0859375,11.2159295133508)(0.09375,10.3668178769203)(0.1015625,9.6419023208275)(0.109375,8.99279219279653)(0.1171875,8.42932340191475)(0.125,7.9192908856329)(0.1328125,7.48514186670232)(0.140625,7.0939539348811)(0.1484375,6.7093645524933)(0.15625,6.37994989047613)(0.1640625,6.08084529715769)(0.171875,5.80834876361631)(0.1796875,5.55789109552482)(0.1875,5.32805303483545)(0.1953125,5.11626626426453)(0.203125,4.9203584655669)(0.2109375,4.74029456873785)(0.21875,4.57000578133464)(0.2265625,4.41356105451219)(0.234375,4.26588569912626) 
};

\end{axis}
\draw  [<-][red][dashed,thick] (0.75,2.5) -- (1.2,2.3);
\draw  [-][red][dashed,thin] (0.63,2.4) -- (0.63,0);
\node (max1) at (2.85,2.2) {\textcolor{red}{\footnotesize{$(M/N \approx 0.0234, T^{\ast} \approx24)$}}};
\node (max1) at (0.64,2.48) {\textcolor{red}{\small x}};

\draw  [<-][red][dashed,thick] (1.4,1) -- (2,1.3);
\draw  [-][red][dashed,thin] (1.35,0.85) -- (1.35,0);
\node (max1) at (3.6,1.34) {\textcolor{red}{\footnotesize{$(M/N \approx 0.0703, T^{\ast} \approx 8)$}}};
\node (max1) at (1.365,0.9) {\textcolor{red}{\small x}};
\end{tikzpicture}\label{fig:performance-b}
        \vspace{-1mm}
    \end{subfigure}\\
    \begin{subfigure}{\columnwidth}
    \caption{}
%
%
%
%
\begin{tikzpicture}

\begin{axis}[%
view={0}{0},
width=6.5cm,
height=2.6cm,
scale only axis,
enlarge x limits=0.05,
  try min ticks=5,
  max space between ticks=30pt,
  x tick label style={
    /pgf/number format/.cd,
    fixed,
    precision=2
  },
xmin=0, xmax=0.4,
xlabel={M/N},
xmajorgrids,
ymin=10, ymax=50,
ylabel={$\beta{}^\text{*}$},
ymajorgrids,
yminorgrids,
label style ={font=\small}]
\addplot [
color=blue,line width=1.0pt,
solid,
mark=,
mark options={solid}
]
coordinates{
 (0.0078125,13)(0.015625,13.6)(0.0234375,14.5)(0.03125,15.6)(0.0390625,16.6)(0.046875,18.3)(0.0546875,20.8)(0.0625,25.3)(0.0703125,26.2)(0.078125,26.9)(0.0859375,27.4)(0.09375,27.8)(0.1015625,28.3)(0.109375,28.7)(0.1171875,28.9)(0.125,29.3)(0.1328125,29.6)(0.140625,29.8)(0.1484375,30.1)(0.15625,30.3)(0.1640625,30.5)(0.171875,30.7)(0.1796875,30.8)(0.1875,31.1)(0.1953125,31.3)(0.203125,31.4)(0.2109375,31.6)(0.21875,31.7)(0.2265625,31.7)(0.234375,31.8)(0.2421875,32.1)(0.25,32.2)(0.2578125,32.3)(0.265625,32.5)(0.2734375,32.6)(0.28125,32.6)(0.2890625,32.8)(0.296875,32.8)(0.3046875,32.9)(0.3125,33)(0.3203125,33)(0.328125,33.2)(0.3359375,33.2)(0.34375,33.4)(0.3515625,33.5)(0.359375,33.5)(0.3671875,33.6)(0.375,33.7)(0.3828125,33.8)(0.390625,33.9) 
};

\addplot [
color=black,
dashed,line width=1.0pt,
mark=,
mark options={dashed}
]
coordinates{
 (0.0078125,24.4)(0.015625,31.7)(0.0234375,37.4)(0.03125,38.7)(0.0390625,39.7)(0.046875,40.4)(0.0546875,41)(0.0625,41.5)(0.0703125,41.9)(0.078125,42.3)(0.0859375,42.7)(0.09375,42.9)(0.1015625,42.9)(0.109375,43.4)(0.1171875,43.5)(0.125,43.6)(0.1328125,43.9)(0.140625,44.1)(0.1484375,44.3)(0.15625,44.5)(0.1640625,44.5)(0.171875,44.8)(0.1796875,44.9)(0.1875,45)(0.1953125,45)(0.203125,45)(0.2109375,45.2)(0.21875,45.5)(0.2265625,45.5)(0.234375,45.5)(0.2421875,45.7)(0.25,45.7)(0.2578125,45.7)(0.265625,45.9)(0.2734375,46.1)(0.28125,46.1)(0.2890625,46.2)(0.296875,46.2)(0.3046875,46.3)(0.3125,46.5) 
};

\end{axis}

\draw  [<-][red][dashed,thick] (0.7,1.75) -- (1,1.7);
\draw  [-][red][dashed,thin] (0.63,1.75) -- (0.63,0);
\node (max1) at (2.65,1.7) {\textcolor{red}{\footnotesize{$(M/N \approx 0.0234, \beta^{\ast} \approx 37)$}}};
\node (max1) at (0.64,1.75) {\textcolor{red}{\small x}};

\draw  [<-][red][dashed,thick] (1.5,1) -- (2,0.9);
\draw  [-][red][dashed,thin] (1.35,1) -- (1.35,0);
\node (max1) at (3.65,0.9) {\textcolor{red}{\footnotesize{$(M/N \approx 0.0703, \beta^{\ast} \approx 26)$}}};
\node (max1) at (1.365,1.05) {\textcolor{red}{\small x}};

\end{tikzpicture}\label{fig:performance-c}
        \vspace{-1mm}
    \end{subfigure}
    \caption{\small a) Maximum user resolution probability $P^\ast_\mathrm{R}$, b) maximum expected throughput $T^\ast$, c) optimum average slot degree $\beta^{\ast}$, as functions of $M/N$, at $1/\sigma^2_\mathrm{n} = 5 \text{ and } 10$ dB.}
 \label{fig:performance}
 \vspace{-6mm}
\end{figure}
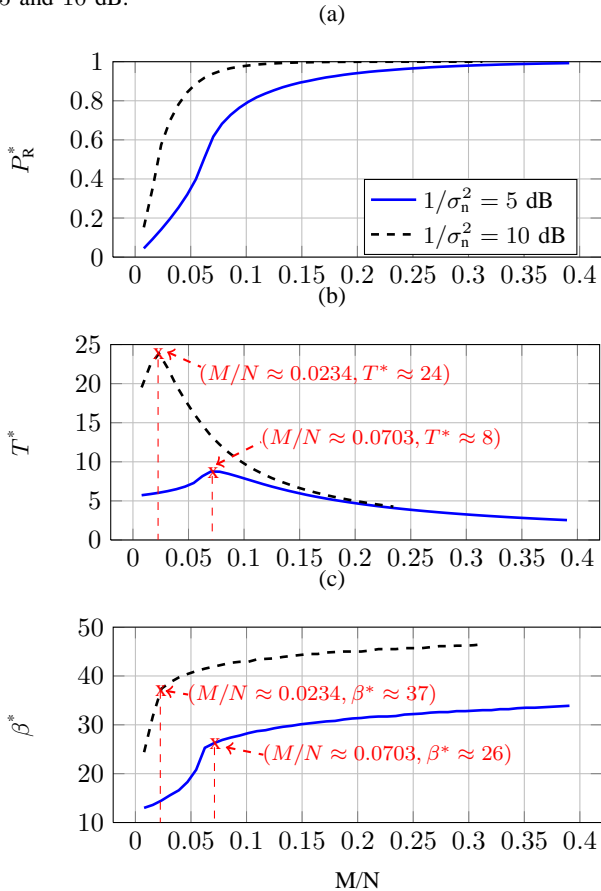

\section{Asymptotic Performance}
\label{sec:performance}

In this section we present asymptotic performance results obtained using and-or tree evaluation with numerically evaluated capture probabilities.
Specifically, we present the maximum user resolution probability $P^\ast_\mathrm{R}$, the corresponding maximum expected throughput $T^\ast$ and the corresponding optimum average slot degree $\beta^{\ast}$ for which $P^\ast_\mathrm{R}$ and $T^\ast$ are obtained, as functions of the ratio of the numbers of users and slots $M/N$. 

Fig.~\ref{fig:performance} shows the asymptotic performance for $1/\sigma^2_\mathrm{n} = 5 \text{ and } 10$~dB; obviously, the trends are the same for both values.
However, the increase of $1/\sigma^2_\mathrm{n}$ highly affects the performance due to the higher capture probabilities.
As $M/N$ increases, $P^\ast_\mathrm{R}$ steeply increases at first and then saturates; this behavior is analogous to the typical behavior of iterative BP erasure decoding \cite{M2005}.
Correspondingly, $T^\ast$ at first increases and then, after some critical $M/N$, starts to drop.
This critical value of $M/N$ actually defines the asymptotically optimal length of the contention period, for which the overall maximum expected throughput can be achieved;
Fig.~\ref{fig:performance}c) shows which $\beta^{\ast}$ should be used for the given $M/N$, in order to achieve this overall maximum expected throughput.
For example, if one chooses $M \approx 0.023 N$ and $\beta^{\ast} \approx 37$ when $1/\sigma^2_\mathrm{n} = 10$~dB, then the expected throughput is $T^\ast \approx 24$.
The critical value of $M/N$ is higher for lower $1/\sigma^2_\mathrm{n}$, due to slower increase in $P^\ast_\mathrm{R}$.

A closer inspection of the results reveals the following.
For $1/\sigma^2_\mathrm{n} = 10$~dB, using \eqref{eq:avg_user_deg} it can be shown that for the critical $M/N$ the expected user degree, i.e., the average number of transmitted replicas per user, is only about 0.87.
This actually means that there are not enough replicas to exploit inter-slot IC and that most of the throughput gain is achieved through intra-slot IC.
On the other hand, for $1/\sigma^2_\mathrm{n} = 5$~dB, the expected user degree is about 1.83 for the critical $M/N$, implying that both inter- and intra-slot IC contribute to throughput performance. 

Finally, the optimum average slot degrees $\beta^{\ast}$ are significantly higher in comparison to scenarios with no capture or narrowband capture effect \cite{SPV2012, ALOHACAP}.
This could be expected, as MUD receivers generally favor higher degrees of collision.

\section{Conclusion}
\label{sec:conclusion}
In this paper we presented a study of the coded slotted ALOHA (SA) with capture effect combined with a a broadband Compressive Sensing based Multi-user Detection (CS-MUD) scheme. This novel access method for coded SA with iterative interference cancellation (IC) exhibits a complex interaction of MAC and PHY layer processing. We analyze the scheme using and-or tree evaluation and the numerically obtained capture probabilities of CS-MUD. The results show that the proposed access method significantly improves the throughput performance by increasing the number of decoded users per slot for coded slotted ALOHA.

\section*{Acknowledgement}

The research presented in this paper was performed in the framework of the FP7 project ICT-317669 METIS, which is partly funded by the European Union. The authors would like to acknowledge the contributions of their colleagues in METIS, although the views expressed are those of the authors and do not necessarily represent the project.


\begin{thebibliography}{10}
\providecommand{\url}[1]{#1}
\csname url@samestyle\endcsname
\providecommand{\newblock}{\relax}
\providecommand{\bibinfo}[2]{#2}
\providecommand{\BIBentrySTDinterwordspacing}{\spaceskip=0pt\relax}
\providecommand{\BIBentryALTinterwordstretchfactor}{4}
\providecommand{\BIBentryALTinterwordspacing}{\spaceskip=\fontdimen2\font plus
\BIBentryALTinterwordstretchfactor\fontdimen3\font minus
  \fontdimen4\font\relax}
\providecommand{\BIBforeignlanguage}[2]{{%
\expandafter\ifx\csname l@#1\endcsname\relax
\typeout{** WARNING: IEEEtran.bst: No hyphenation pattern has been}%
\typeout{** loaded for the language `#1'. Using the pattern for}%
\typeout{** the default language instead.}%
\else
\language=\csname l@#1\endcsname
\fi
#2}}
\providecommand{\BIBdecl}{\relax}
\BIBdecl

\bibitem{CGH2007}
E.~Cassini, R.~D. Gaudenzi, and O.~del Rio~Herrero, ``{C}ontention {R}esolution
  {D}iversity {S}lotted {ALOHA} {(CRDSA)}: {A}n {E}nhanced {R}andom {A}ccess
  {S}cheme for {S}atellite {A}ccess {P}acket {N}etworks,'' \emph{IEEE Trans.
  Wireless Commun.}, vol.~6, no.~4, pp. 1408--1419, Apr. 2007.

\bibitem{L2011}
G.~Liva, ``{G}raph-{B}ased {A}nalysis and {O}ptimization of {C}ontention
  {R}esolution {D}iversity {S}lotted {ALOHA},'' \emph{IEEE Trans. Commun.},
  vol.~59, no.~2, pp. 477--487, Feb. 2011.

\bibitem{SP2013}
C.~Stefanovic and P.~Popovski, ``{ALOHA} {R}andom {A}ccess that {O}perates as a
  {R}ateless {C}ode,'' \emph{IEEE Trans. Commun.}, vol.~61, no.~11, pp.
  4653--4662, Nov. 2013.

\bibitem{LMS1998}
M.~G. Luby, M.~Mitzenmacher, and A.~Shokrollahi, ``{A}nalysis of {R}andom
  {P}rocesses via {A}nd-{O}r {T}ree {E}valuation,'' in \emph{Proc. of 9th
  ACM-SIAM SODA}, San Francisco, CA, USA, Jan. 1998.

\bibitem{ALOHACAP}
C.~Stefanovic, M.~Momoda, and P.~Popovski, ``{E}xploiting {C}apture {E}ffect in
  {F}rameless {ALOHA} for {M}assive {W}ireless {R}andom {A}ccess,'' in
  \emph{Proc. of IEEE WCNC'14}, Istanbul, Turkey, Apr. 2014.

\bibitem{CS-MUD}
C.~Bockelmann, H.~Schepker, and A.~Dekorsy, ``Compressive sensing based
  multi-user detection for machine-to-machine communication,'' \emph{Trans. on
  ETT}, vol.~24, no.~4, pp. 389--400, Jun 2013.

\bibitem{ISWCS11}
H.~F. Schepker and A.~Dekorsy, ``Sparse multi-user detection for {CDMA}
  transmission using greedy algorithms,'' in \emph{8th ISWCS}, Aachen, Germany,
  November 2011.

\bibitem{VTC12}
H.~Schepker and A.~Dekorsy, ``{C}ompressive {S}ensing {M}ulti-{U}ser
  {D}etection with {B}lock-{W}ise {O}rthogonal {L}east {S}quares,'' in
  \emph{IEEE 75th VTC}, Yokohama, Japan, May 2012.

\bibitem{SPV2012}
C.~Stefanovic, P.~Popovski, and D.~Vukobratovic, ``{F}rameless {ALOHA} protocol
  for {W}ireless {Networks},'' \emph{IEEE Comm. Letters}, vol.~16, no.~12, pp.
  2087--2090, Dec. 2012.

\bibitem{NP2012}
K.~R. Narayanan and H.~D. Pfister, ``Iterative {Co}llision {R}esolution for
  {S}lotted {ALOHA}: {A}n {O}ptimal {U}ncoordinated {T}ransmission {P}olicy,''
  in \emph{Proc. of 7th ISTC}, Gothenburg, Sweden, Aug. 2012.

\bibitem{R1975}
L.~G. Roberts, ``{ALOHA} packet system with and without slots and capture,''
  \emph{SIGCOMM Comput. Commun. Rev.}, vol.~5, no.~2, pp. 28--42, Apr 1975.

\bibitem{N1984}
C.~Namislo, ``{A}nalysis of {M}obile {R}adio {S}lotted {ALOHA} {N}etworks,''
  \emph{IEEE J. Sel. Areas Commun.}, vol.~2, no.~4, pp. 583--588, Jul. 1984.

\bibitem{GVS1988}
S.~Ghez, S.~Verdu, and S.~C. Schwartz, ``{S}tability {P}roperties of {S}lotted
  {ALOHA} with {M}ultipacket {R}eception {C}apability,'' \emph{IEEE Trans.
  Autom. Control}, vol.~33, no.~7, pp. 640--649, Jul 1988.

\bibitem{ZR1994}
M.~Zorzi and R.~R. Rao, ``{C}apture and {R}etransmission {C}ontrol in {M}obile
  {R}adio,'' \emph{IEEE J. Sel. Areas Commun.}, vol.~2, no.~4, pp. 1289--1298,
  Oct. 1994.

\bibitem{NEW2007}
G.~D. Nguyen, A.~Ephremides, and J.~E. Wieselthier, ``{O}n {C}apture in
  {R}andom-{A}ccess {S}ystems,'' in \emph{Proc. 2006 IEEE Int. Symp. Inf.
  Theory}, Seattle, WA, USA, Jul. 2006.

\bibitem{TheoreticalCapture}
A.~Zanella and M.~Zorzi, ``{T}heoretical {A}nalysis of the {C}apture
  {P}robability in {W}ireless {S}ystems with {M}ultiple {P}acket {R}eception
  {C}apabilities,'' \emph{IEEE Trans. Commun.}, vol.~60, no.~4, pp. 1058--1071,
  April 2012.

\bibitem{Verdu}
S.~Verd{\'u}, \emph{Multiuser Detection}.\hskip 1em plus 0.5em minus
  0.4em\relax Cambridge, U.K.: Cambridge Univ. Press, November 1998.

\bibitem{RU2007}
T.~Richardson and R.~Urbanke, \emph{{M}odern {C}oding {T}heory}.\hskip 1em plus
  0.5em minus 0.4em\relax Cambridge University Press, Cambridge, UK, 2007.

\bibitem{M2005}
D.~J.~C. MacKay, ``{F}ountain {C}odes,'' \emph{IEE Proceedings on
  Communications}, vol. 152, no.~6, pp. 1062--1068, 2005.

\end{thebibliography}
\end{document}